\documentclass[paper]{aa}

\usepackage{graphicx}

\begin{document}

\def\P{\bar{\Phi}}

\def\st{\sigma_{\rm T}}

\def\vk{v_{\rm K}}

\def\sles{\lower2pt\hbox{$\buildrel {\scriptstyle <}
   \over {\scriptstyle\sim}$}}

\def\sgreat{\lower2pt\hbox{$\buildrel {\scriptstyle >}
   \over {\scriptstyle\sim}$}}

\title{A note on the cyclic evolution of the pulsar magnetosphere}

\author{Ioannis Contopoulos}
\institute{Research Center for Astronomy, Academy of Athens, 
GR-11527 Athens, Greece\\
\email{icontop@academyofathens.gr}}

\titlerunning{Cyclic pulsar spindown}

\date{Received / Accepted }

\abstract
{

Positive and negative pulsar breaking indices
suggest that some fraction of the pulsar
spindown torque undergoes a cyclic evolution.
The observed  strong correlation of
`anomalous' breaking indices with pulsar age
implies that the characteristic periodicity
timescale is in the range
100 to 10,000 years depending on the
fraction of the spindown torque that undergoes
cyclic evolution, 1 to 100\% respectively. 
We argue that the longest variability timescale 
is consistent with a neutron star magnetic cycle
similar to the solar cycle.


}

\maketitle

\section{Observational evidence}

A pulsar spins down due to the torque 
on the neutron star crust generated by the electric current
flowing in its magnetosphere. In the simplified picture of a steady-state
axisymmetric force-free
ideal MHD magnetosphere, Contopoulos, Kazanas \& Fendt 1999 
(hereafter CKF) first showed
that the distribution of the magnetospheric electric current 
$I$ can be determined as an eigenvalue of the problem, 
if one makes the natural assumption that the
magnetosphere is smooth and continuous on the light cylinder
(defined as the distance $r_L\equiv c/\Omega$ from the rotation
axis, where $\Omega$ is the pulsar angular velocity).
The unique electric current distribution thus obtained yields a
unique pulsar spindown torque, and thus a unique pulsar
spindown rate $\dot{\Omega}$. 
This is of the same order as the value obtained
for simple electromagnetic vacuum dipole radiation, namely
\begin{equation}
\dot{\Omega}=-f\frac{B_*^2r_*^4}{c^3 M_*}\Omega^3\ .
\label{Omegadot}
\end{equation}
Here, $B_*$, $r_*$, and $M_*$ are the neutron star
polar magnetic field, radius and mass respectively; and $f$ is
a numerical factor of order unity. It turns out that
this result is also valid in the general non-axisymmetric
case (Bogovalov~1998; Spitkovsky~2006; Contopoulos~2007a), and
therefore, one may use Eq.~\ref{Omegadot} to obtain an
estimate of the polar value of the neutron star magnetic field
from measurements of $\Omega$ and $\dot{\Omega}$.

Strictly speaking, however, the results of CKF and subsequent 
related work 
are only valid in steady-state. Thus, as the neutron star
spins down and the light cylinder moves to larger and larger
distances, one needs to take into account the
evolution of the pulsar magnetosphere. The first thing one
may assume is that the magnetosphere evolves through a
sequence of steady-state equilibria of the CKF type, i.e. that
it manages to readjust itself so that
at all times, the region of closed lines extends all the
way to the light cylinder, and the last open magnetic field line
extends to infinite distances without reconnecting accross the
equator. In addition, one may assume that $B_*$
does not evolve with pulsar age.
Unfortunately, the situation is more complicated
than that, since Eq.~\ref{Omegadot} yields a braking index value
\begin{equation}
n\equiv \frac{\ddot{\Omega}\Omega}{\dot{\Omega}^2} = 3\ ,
\label{nsimple}
\end{equation}
and most known measurements of $n$ differ from that value.
In fact, there exist today $\sim 400$ pulsars in the ATNF catalogue
(Manchester {\em et al.}~2005) with measured values of $n$
in the range from $-10^6$ to $+10^6$. Although all but six of these
values are characterized in the literature as `anomalous', one
thing is certain: Eq.~\ref{nsimple} cannot be right in general.

In Figure~1 we plot
$\pm \log|n|$ (with $\pm$ according to whether $n>0$ or $n<0$
respectively) as a function of the characteristic spindown time
$\tau\equiv -\Omega/(2\dot{\Omega})$ in years. One may argue
(Alice Harding, personal communication) that in
young pulsars ($\tau < 10^5$), braking index measurements
may be `corrupted' by neutron star glitches. On the other hand, in older
pulsars ($\tau > 10^5$) where glitches are not as important,
one finds the correlation
\begin{equation}
\pm \log|n| \sim \pm ( \log\tau - 3 )\ .
\label{nobs}
\end{equation}
On the other hand, in a $P-\dot{P}$ diagram ($P\equiv 2\pi/\Omega$ 
is the pulsar period) containing
all the cases with observed values of $n$, there is
no obvious correlation between $P$ and $\dot{P}$ (Figure~2).
In other words, the pulsars with measured values of $n$ 
have been taken randomly
from the total number of known pulsars without any obvious observational
bias, and therefore, they represent pulsars at various stages
of their evolution. 
Beskin, Biryukov \& Karpov~(2006) argued that
the $\pm$ symmetry of Figure~1 implies that some fraction 
of the pulsar angular velocity undergoes a cyclic evolution.
We suggest that it is more constructive to consider a cyclic component
in the evolution of the pulsar spindown torque (which is, after all, the
source of the angular velocity evolution).
Interestingly enough, as we will see next, the data
also yields some information on the periodicity timescale.

There are several ways to reconcile eqs.~\ref{Omegadot} and \ref{nobs}:
(A) Assume that the approximation of a sequence of
CKF-type steady-state magnetospheric equilibria holds,
and that the neutron star magnetic field undergoes a cyclic evolution;
(B) Relax the assumptions of the CKF analysis and assume
a variable magnetospheric
structure that would yield a cyclic evolution of the factor $f$;
(C) Relax the assumption of constant neutron star
moment of inertia. Only (A) and (B) refer to the spindown
torque itself. In any case,
\begin{equation}
\dot{\Omega}  = -f_o\frac{B_{*o}^2 r_*^4}{c^3 M_*}\Omega^3 F(t)\ ,
\label{Omegadot2}
\end{equation}
with $f_o$, $B_{*o}$ characteristic values of the spindown
parameter $f$ and the polar magnetic field $B_*$ respectively, and
\begin{equation}
F(t) =
1-\frac{\alpha}{2}+\frac{\alpha}{2}\cos(2\pi\frac{t}{\tau_{cycle}}+\phi)
\label{F}
\end{equation}
characterizing the cyclic variation of either $f$ or $B_*$.
Here, $\tau_{cycle}$ is the characteristic period of the cyclic
spindown evolution in years; 
$\alpha$ is the fraction of the spindown torque that
varies periodically ($0\leq \alpha \leq 100\%$);
and $\phi$ is a random initial (at pulsar birth)
phase angle. Note that, in order for the star to continuously 
spin down we must have $F\geq 0$ at all times. Therefore,
\begin{equation}
n  =  3+\frac{\Omega/\dot{\Omega}}{F/\dot{F}}
\sim 3+2\pi \alpha \frac{\tau}{\tau_{cycle}}\ .
\label{ncyclic}
\end{equation}
For old pulsars with $\tau \sgreat \, \tau_{cycle}$, Eq.~\ref{ncyclic}
can equivalently be written as
\begin{equation}
\pm \log|n| \sim \pm ( \log\tau - \log\tau_{cycle} + 1 
+ \log\alpha )\ .
\label{ncyclic2}
\end{equation}
Comparing eqs.~\ref{nobs} and \ref{ncyclic2},
one obtains the following approximate relation between the
characteristic period and the fraction of
the pulsar spindown which varies in a cyclic way,
\begin{equation}
\log\tau_{cycle} \sim 4 + \log\alpha\ .
\label{cycle}
\end{equation}
We plot in Figure~3 what Eq.~\ref{ncyclic} yields 
for the $\sim 400$ pulsars
with measured values of $n$, assuming $\alpha=100\%$.
Note that the fit is independent of $f_o$ and $B_{*o}$. 
As Beskin, Biryukov \& Karpov~(2006) suggested,
the minimum characteristic period cannot be smaller
than the pulsar observation period of 40~years, 
and this together with Eq.~\ref{cycle} yields a range
$100 \, \sles \, \tau_{cycle} \, \sles \, 10,000$ for
$1\% \leq \alpha \leq 100\%$ respectively.


\section{Cyclic magnetospheric evolution}

Several physical models 
that address the issue of cyclic variation of
the pulsar spindown
have been proposed in the literature, ranging
from neutron star interior ``wobbling''
on a timescale of a few years (e.g. Kundt~1988), to
magnetospheric variability (e.g. Contopoulos~2005).
In the present work, we would like to focus on
our simplest (one-parameter) fit of the anomalous braking index 
data, namely the one with $\tau_{cycle}\sim 10,000$
and $\alpha\approx 100\%$. 

$F(t)$ becoming zero periodically
is not compatible with a cyclic
evolution of the neutron star moment of inertia
(case C above). On the other hand
such a scenario is compatible with a cyclic evolution of the neutron
star magnetic field similar to the solar cycle (case A above).
Interestingly enough,
the ten thousand year timescale that we obtain
is comparable to the neutron star
cooling timescale (e.g. Blandford, Applegate \& Hernquist~1983).
It is conceivable that some sort of
dynamo mechanism in the neutron star interior, may support
a cyclic evolution with
\begin{equation}
B_* = B_{*o}\sin(2\pi\frac{t}{\mbox{5,000\ years}}+\phi)
\label{Bstar}
\end{equation}
(Eqs.~\ref{Omegadot2}, \ref{F}). 
Note that this scenario does not require
magnetic field decay (at least over
timescales shorter than about $10^{11}$~years), in
agreement with the analysis of the $P-\dot{P}$ diagram presented
in Contopoulos \& Spitkovsky~(2006).

We also tried to seek variants of the CKF solution (case B above)
that would yield values of $f$ very different from unity.
In fact, what we need is a physical
mechanism that will periodically turn off
the neutron star magnetospheric spindown.
In a series of papers (Contopoulos~2007b,c), we relaxed
the assumption of ideal MHD in the equatorial
region of the pulsar magnetosphere beyond the light cylinder.
This is the region where the magnetospheric return current
flows, and several authors before us suggested that this
may be the region of electromagnetic energy dissipation
that would result in particle acceleration (e.g.
Coroniti~1990; Michel~1994; Lyubarsky \& Kirk~2001; 
Kirk \& Skj\ae raasen~2003; Romanova, Chulsky \& Lovelace~2005).
As we argued in Contopoulos~2007c, one cannot study equatorial
reconnection without taking into account the global topology
of the poloidal magnetic field.
The details of equatorial reconnection remain (yet) unknown.
However, it is easy to realize that, when equatorial reconnection
is present, magnetic field lines that cross the light cylinder and would
have extended to infinity in CKF, now continuously reconnect
across the equator. As a result, 
the equatorial condition for the magnetic flux
function $\Psi(r;z)$ 
(defined as the magnetic flux
crossing a circle of cylindrical radius $r$
at height $z$ around the axis of rotation)
differs from that in CKF. In particular, 
$\Psi(r>r_L;z=0)$ is not constant but decreases
with distance. We assume for simplicity that
\begin{equation}
\Psi(r>r_L;z=0)=\Psi(r=r_L;z=0)(r/r_L)^{-\epsilon}\ ,
\label{Psieq}
\end{equation}
where, $\epsilon$ is a parameter that characterizes the
effect of dissipation ($\epsilon=0$ corresponds to the ideal MHD
case studied in CKF, whereas $\epsilon=1$ corresponds
to a magnetosphere with maximum equatorial dissipation).
Equation~\ref{Psieq} is a new (to our knowledge) equatorial
boundary condition beyond the light cylinder, and one may
thus implement the same procedure as described in CKF to
solve the pulsar equation (Scharlemann \& Wagoner~1973), and thus
obtain the magnetospheric structure and electric current distribution
$I(\Psi)$ for various values of $0\leq \epsilon \leq 1$, 
as seen in Figures~4-6. 
For each such electric current distribution the
total electromagnetic spindown torque acting on the neutron-star
crust is proportional to the integral $\int I(\Psi){\rm d}\Psi$ 
(e.g. Michel~1991). 
Note that when $\epsilon=0.4$ (Fig.~5) there
is no equatorial return current sheet (the return current is
distributed along the magnetic field lines that cross the
light cylinder), whereas when $\epsilon=1$ (Fig.~6)
$\int I(\Psi){\rm d}\Psi \sim 0$, 
i.e. the total neutron-star spindown torque is close to zero.
In Figure~7 we 
plot the value of the spindown torque parameter $f$ as a function 
of our dissipation parameter $\epsilon$. One sees that,
as we introduce more and more dissipation in the
equatorial region, the magnetosphere evolves to a configuration
with less and less electromagnetic torque acting on the
central neutron star. 
Obviously, a cyclic evolution of the physical mechanism that
allows or inhibits equatorial dissipation in the pulsar magnetosphere
(e.g. variability in the supply
of charge carriers from the neutron star surface that may be
due to a periodic stellar wind)
would yield a cyclic evolution of the magnetospheric
torque. Unfortunately, magnetospheric solutions 
for $0.4\leq\epsilon\leq 1$ 
($f\leq 0.5$, or equivalently
$F\leq 80\%$) contain regions with $I(\Psi)<0$,
where electromagnetic energy is flowing {\em from} the
magnetosphere {\em onto} the star, and therefore, 
such solutions are probably unphysical.

We conclude that a cyclic component in the evolution
of the magnetospheric spindown torque
may account for the measured large positive and negative
anomalous braking index values. 
If we are willing to consider a 100\% cyclic evolution,
this can only be due to a neutron star magnetic 
cycle similar to the solar cycle.
In that case, 
the evolution timescale would be on the order of 10,000~years.

\acknowledgements{We would like to thank Alice Harding and 
Demos Kazanas for
their hospitality at the NASA Goddard Space Flight Center in
January and June 2007
where some of the ideas in the present work originated.
We would also like to thank Pr. Wolfgang Kundt
for an honest exchange of ideas.}

\newpage
\begin{figure}
\includegraphics[angle=270,scale=.40]{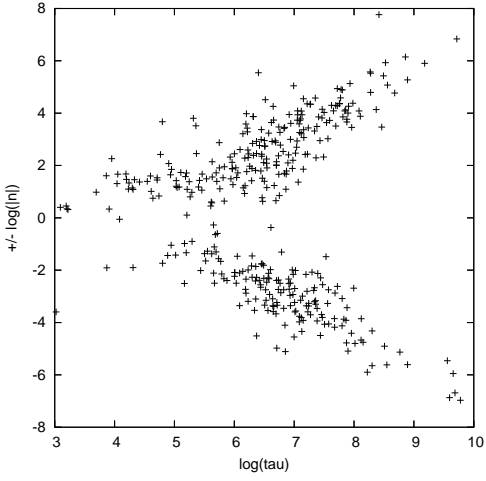}
\caption{Braking index as a function of characteristic
spindown time. We plot here $\pm \log|n|$ ($\pm$ according
to whether $n>0$ or $n<0$) vs. $\log\tau$, where
$\tau\equiv -\Omega/2\dot{\Omega}$ in years.
Note that $|n|>1$ everywhere.
At large $\tau >10^5$, 
the diagram may be fit by the simple linear relation
$\pm \log|n|\sim \pm (\log\tau-3 )$.
}
\label{fig1}
\end{figure}

\begin{figure}
\includegraphics[angle=270,scale=.40]{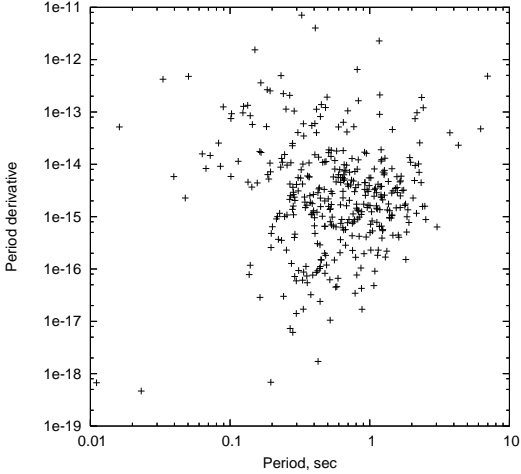}
\caption{$P-\dot{P}$ diagram for the $\sim 400$
pulsars with measured braking index values shown
in Figure~1. Here, $P$, $\dot{P}$ are the pulsar
period and period derivative respectively.
The distribution is that of a standard
sparse $P-\dot{P}$ diagram without any obvious
observational bias.
}
\label{fig2}
\end{figure}

\begin{figure}
\includegraphics[angle=270,scale=.40]{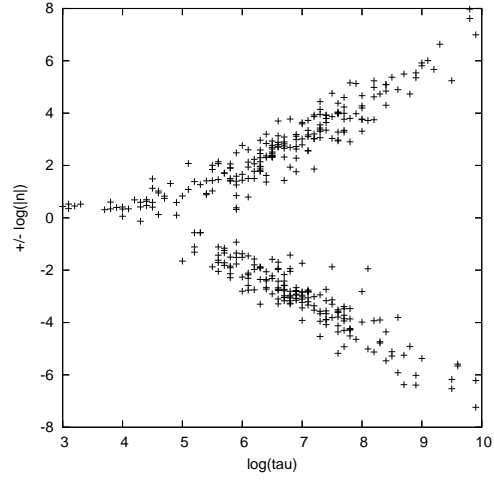}
\caption{Fit of the distribution shown in Figure~1 assuming 
a 100\% cyclic evolution of the pulsar spindown torque 
(Eqs.~\ref{Omegadot2}
\& \ref{F}). $\tau_{cycle}\sim 10,000$ (in years).
The fit is acceptable, even for young pulsars
($\tau\leq 10^5$) where some of the dispersion in
the measurements of $n$ is due to neutron star glitches.
}
\label{fig3}
\end{figure}


\begin{figure}
\includegraphics[angle=270,scale=.40]{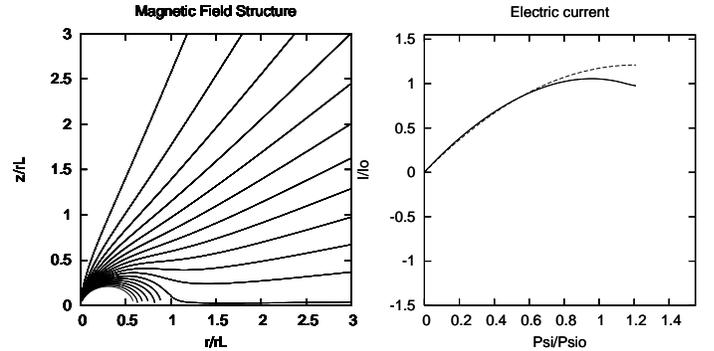}
\caption{On the left, we plot the magnetic field structure
in the case of no magnetospheric reconnection (CKF).
Distances are normalized to the light cylinder distance $r_L$.
On the right, we plot the corresponding electric current distribution
$I=I(\Psi)$
along the field lines that cross the light cylinder.
The magnetic flux is normalized to the canonical value
$\Psi_o\equiv \pi B_* r^3/r_L$.  The electric current is
normalized to the canonical value $I_o\equiv \Omega \Psi_o/(4\pi)$.
For comparison, we plot also (dashed line)
the electric current distribution 
of a relativistic magnetic split monopole with the same amount
of magnetic flux crossing the light cylinder (Michel~1991).
}
\label{fig5}
\end{figure}

\begin{figure}
\includegraphics[angle=270,scale=.40]{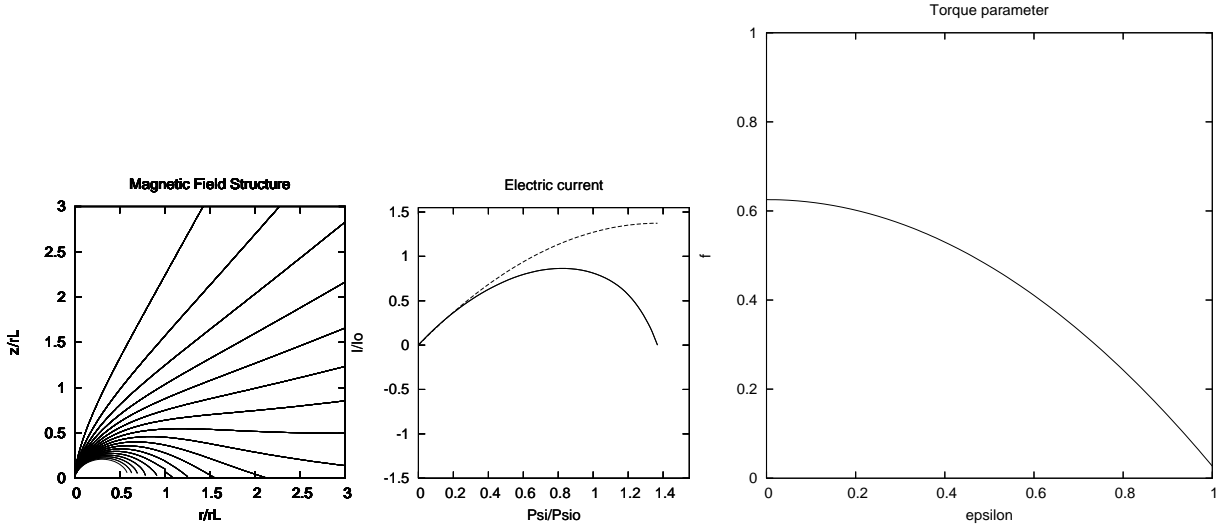}
\caption{Same as Fig.~4 with some amount of equatorial magnetospheric
reconnection that corresponds to 
$\epsilon = 0.4$ (Eq.~\ref{Psieq}). 
}
\label{fig6}
\end{figure}

\begin{figure}
\includegraphics[angle=270,scale=.40]{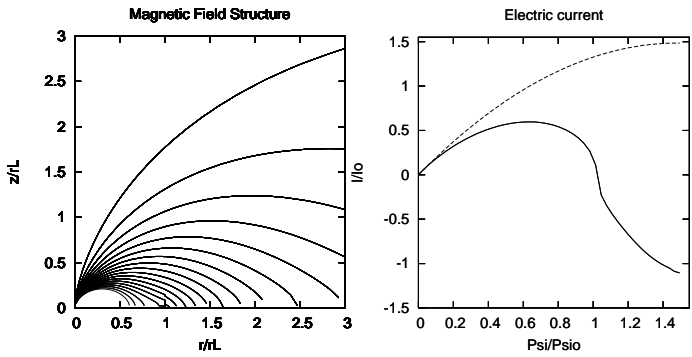}
\caption{Same as Fig.~4 with maximum equatorial magnetospheric
reconnection that corresponds to
$\epsilon = 1$. 
}
\label{fig7}
\end{figure}

\begin{figure}
\includegraphics[angle=270,scale=.40]{8108fig7.ps}
\caption{The spindown torque parameter $f$ (Eq.~\ref{Omegadot})
as a function of the dissipation parameter $\epsilon$. 
In the absence of reconnection ($\epsilon = 0$), $f=0.6$ (CKF).
}
\label{fig8}
\end{figure}

\end{document}